\documentclass[11pt]{article} 
\usepackage[dvips]{graphics}
\usepackage  {epsfig}                                                         
\newfont{\rams}{msbm10 scaled\magstep1}
\newfont{\fams}{msbm10}
\newfont{\iams}{msbm8}
\newfont{\gotic}{eufm10 scaled\magstep1}
\newfont{\bellap}{eusm10 scaled\magstep1}
\newfont{\bellaps}{eusm7} 
\newcommand{\rea}{\mbox{\rams \symbol{'122}}}

\newcommand{\rind}{\mbox{\iams \symbol{'122}}} 
\newcommand{\ecar}{\mbox{\gotic \symbol{'145}}}

\newcommand{\Kb}{\hspace{1pt} \mbox{\bellap K}}
\newcommand{\Kbop}{\Kb_{op}}
\newcommand{\Kbac}{\Kb_{ac}}

\newcommand{\nq}{{\it n}_{op}}
\newcommand{\hw}{\hbar \omega_{op}}
\newcommand{\kB}{k_{B}}
\newcommand{\TL}{T_{L}}
\newcommand{\kT}{\kB \TL}
\newcommand{\mass}{m^{*}}

\newcommand{\dm}{\displaystyle}
\newcommand{\p}{\hspace{1pt} .}
\newcommand{\sv}{\hspace{1pt} ,}
\newcommand{\ud}{:=}
\newcommand{\freccia}{\rightarrow}

\newcommand{\bE}{{\bf E}}
\newcommand{\bk}{{\bf k}}

\newcommand{\bkt}{\tilde{{\bf k}}}
\newcommand{\bx}{{\bf x}}

\newcommand{\txk}{(t, \bx, \bk)}
\newcommand{\vk}{{\bf v}(\bk)}
\newcommand{\ftxk}{f \txk}

\newcommand{\fk}{f (\bk)}

\newcommand{\fbo}{f_{0}(\en)}
\newcommand{\fbi}{f_{1}(\en)}

\newcommand{\su}[1]{\sigma(#1)}

\newcommand{\en}{\varepsilon}

\newcommand{\enk}{\en(\bk)}

\newcommand{\enkt}{\en(\bkt)}

\newcommand{\Rk}{ \rea^{3}}

\newcommand{\devp}[2]{ \frac{\partial #1}{\partial #2}}
\newcommand{\devo}[2]{ \frac{d \hspace{1pt} #1}{d \hspace{1pt} #2}}
\newcommand{\Itre}{\int}
\newcommand{\Itrek}{\int_{\scriptstyle \rind^{3}}}

\newcommand{\Ik}[1]{\Itrek #1 \hspace{2pt} d \bk}

\newcommand{\Ikt}[1]{\Itre #1 \hspace{2pt} d \bkt}
\newcommand{\Ikb}[1]{\Itre #1 \hspace{2pt} d \bk}

\newcommand{\Ie}[1]{\int_{0}^{+ \infty} #1 \hspace{2pt} d\en}

\newcommand{\ddm}{\hspace{1pt} \delta (\en - \tilde{\en} - \hw) }
\newcommand{\ddp}{\hspace{1pt} \delta (\en - \tilde{\en} + \hw) }
\newcommand{\dee}{\hspace{1pt} \delta (\en - \tilde{\en}) } 
\newcommand{\deet}{\hspace{1pt} \delta (\enk - \tilde{\en}) }

\newcommand{\eq}[1]{\mbox{{\rm(\ref{#1})}}}
\begin{document}
\pagestyle{plain}
%
%
%
\begin{center}
{\Large \bf An asymptotic solution for the SHE equations describing 
the charge transport in semiconductors.}
\\[5mm]
{\Large \sc Salvatore Fabio Liotta} 
\\
{Dipartimento di Matematica e Informatica,} 
{Universit\`{a} di Catania  } \\
{Viale A.~Doria 6, 95125, Catania, Italy}
\end{center}
{\bf Abstract}  In this paper an asymptotic solution of 
the spherical harmonics equations describing the charge transport 
in semiconductors is found. This solution is compared with a 
numerical solution for bulk silicon device. We also indicate 
application of this solution to the construction of 
high field hydrodynamical models.\\[5pt]
\section{Introduction.}
In the framework of charge transport in semiconductors, a technique
widely used in order to find approximate solutions of the 
Boltzmann transport equation (BTE) is based
on a spherical harmonics expansion (SHE) of the distribution 
function (Rahmat, White and Antoniadis,~1996; Vecchi and Rudan,~1998;
Ventura, Gnudi and Baccarani,~1995; Liotta and Struchtrup,~2000). 
Recently an asymptotic solution of the SHE
equations was found (Liotta and Majorana,~1999) 
in the case of a homogeneous 
(bulk) device with a simple parabolic band structure. 
Despite the very simple situation in which this solution was obtained, 
it has revealed to be very useful in order to develop new asymptotic
hydrodynamical models describing the hot electron population 
in silicon devices (both in the homogeneous and non-homogeneous case).
In particular see Anile and Mascali~(2000) and Anile, Liotta and Mascali~(2000),
where this asymptotic 
solution was used in order to close the set of moment equations.\\
The aim of this work is to show the possibility of finding a new 
asymptotic solution generalizing that derived in Liotta and Majorana~(1999)
to the case of a non-parabolic band structure (Kane model). 
this solution reduces to the old one when
the non-parabolicity parameter goes to zero. 
The importance is due to the fact that the Kane equation fits better 
the real band structure in the high field regime.
Therefore this solution can be very useful in order to develop 
improved high field hydrodynamical models which could 
describe better the hot electron population. This solution is also 
interesting by itself in the framework of SHE models. 
\section{Basic equations.}
We consider the case of unipolar semiconductor devices in which 
the current is essentially due to electrons 
(but the results can be generalized to holes). The semiclassical
description of the electron transport is based on the BTE 
(Markowich et al.,~1990; Ferry,~1991; Cercignani,~1987),
which writes
\begin{equation}
 \devp{f}{t} +\vk \cdot \nabla_{\bx} f -
 \frac{\ecar}{\hbar} \bE \cdot \nabla_{\bk} f = Q(f) \sv
 \label{BTE} 
\end{equation}
here $\ftxk$ is the electron distribution function, generally depending 
on time $t$, position $\bx$ and wave vector $\bk$ (belonging to the 
first Brillouin zone $B$). $\ecar$ is the 
absolute value of the electron charge, $\hbar$ the reduced 
Planck constant, $\bE$ the electric field. 
$\nabla_{\bx}$ and $\nabla_{\bk}$ denote the gradient with respect 
to $\bx$ and $\bk$ respectively. The group velocity $\vk$ is determined 
by the conduction band structure: $\vk = \frac{1}{\hbar} \nabla_{\bk} \enk$, 
where $\enk$ is the electron energy which depends on the wave vector.
$Q$ is the collision operator which in the non-degenerate case has the form
\begin{equation}
Q(f) =  \Ikt{W(\bk,\bkt) f(t,\bx,\bkt)} 
 - \ftxk \Ikt{W(\bkt,\bk)}  \label{collop}   \sv
\end{equation}
$W(\bk,\bkt)$ representing the electron scattering rate from a 
state with wave vector $\bkt$ to one with wave vector $\bk$.\\
We will consider a stationary and homogeneous situation (bulk device), 
neglecting the Poisson equation and taking into account 
only a constant externally applied electric field, then 
it will be $f=\fk$. Moreover, we will suppose that the electric field is
directed along the $\bx$-axis so to have a cylindrical symmetry around
this axis and represent the two-dimensional momentum space
by means of the polar coordinates $k=|\bk| \left(=\xi(\en)\right)$ and 
$\theta=\arccos{({\bk \cdot \bk_x}/{|\bk||\bk_x|})}$, where
$\bk_x$ is the projection of $\bk$ along $x$. 
Therefore the distribution function can be
expanded in Legendre polynomials of the angle $\theta$
(Rahamat et al.,~1996; Liotta and Struchtrup,~2000) 
\begin{equation}
\fk = \sum_{n} f_{n}(\en) P_{n}(\cos{\theta}) 
\label{svp1} \sv
\end{equation}
where $P_{n}$ is the $n$th order Legendre
polynomial. This expansion will be computationally viable only if few
spherical harmonics are enough to accurately represent the momentum space
distribution. We will assume that the first two terms of the previous 
expression give a good approximation
\begin{equation}
\fk \simeq \fbo +\fbi \cos{\theta} \label{svp2} \p
\end{equation}
The lowest order harmonic coefficient furnishes information about 
the isotropic part of the distribution function and 
$\Ikb{f_{0}}$ yields the electron concentration. 
The first order harmonic coefficient describes the asymmetry of 
the distribution function in the direction of the applied electric field, and $\Ikb{f_{1} \cos{\theta} \vk}/
\Ikb{f_{0}}$ gives the hydrodynamical velocity of the electron gas.\\
We will assume a spherically symmetric band structure of the Kane 
form (Ferry,~1991, Jacoboni and Lugli,~1989; Tomizawa,~1993)
\begin{equation}   
\gamma(\en) \ud \en (1 + \alpha \en) = \frac{\hbar^2 k^2}{2 \mass} \sv
\label{Kane}
\end{equation}
where $\mass$ is the electron effective mass and $\alpha$ the 
constant non-parabolicity parameter. 
By putting $\alpha=0$ one obtains the usual parabolic
band approximation. With this choice we can assume for the first 
Brillouin zone $B \equiv \Rk$ and we have 
$ \vk = \frac{\hbar \bk}{\mass (2 \alpha \en +1)}$.\\
As regards collisions, we will take into account the 
interaction between electrons and
non-polar optical phonons and that between electrons and acoustical phonons,
the latter in the elastic approximation, valid when the thermal energy 
is much greater than that of the phonon involved in the scattering. 
We consider the electron scatterings with ionized impurities
to be negligible, {\it i.e.} we assume the doping density to
be low.
Then the transition rate of the collision operator reads 
(Jacoboni and Lugli,~1989; Tomizawa,~1993)
\begin{equation}
W(\bk,\bkt) = \Kbop \left[\nq~\ddm +(\nq+1)~\ddp \right] + \Kbac \dee 
\label{scatt} \sv
\end{equation} 
where $\en=\enk$, $\tilde{\en}=\enkt$, 
$\nq = \left(\exp{\left(\frac{\hw}{\kB \TL}\right)} -1 \right)^{-1}$ is 
the thermal equilibrium optical phonon number and 
$\Kbop$ and $\Kbac$ are respectively the non-polar
optical and acoustical kernel coefficients 
(constant at a first approximation).
$\hw$ is the optical phonon energy, $\kB$ the Boltzmann constant and $\TL$
the lattice temperature. These choices are appropriate for silicon devices.\\
The SHE equations are easily obtained by inserting the expansion \eq{svp1}
into the BTE \eq{BTE} and balancing the terms of the same order in 
$P_{n}(\cos{\theta})$. To generate a closed set of equations,
all coefficients of order higher than the first are set to be zero,
see Rahmat et al.~(1996) 
(a closure inspired by the Grad moment method, see Grad,~1958).\\ 
But for the aims of this paper it is preferable to perform a change 
of variables and write down
a set of two coupled equations in the unknowns
\begin{eqnarray}
N(\en) = \sigma(\en) f_{0}(\en) \sv \label{N} \\
P(\en) = \frac{8}{3} \pi \frac{\sqrt{\mass}}{\hbar^3} 
\gamma(\en) f_{1}(\en) \sv \label{P}
\end{eqnarray}
where
\begin{equation}
\su{\tilde{\en}} \ud \Ik{\deet} =  \nonumber \\
4 \sqrt{2} \pi \left( \frac{\sqrt{\mass}}{\hbar} 
\right)^{3} H(\tilde{\en}) (\gamma(\tilde{\en}))^{\frac{1}{2}}~ 
\gamma'(\tilde{\en})  \label{dos}
\end{equation}
is the density of states . $H(\en)$ is the Heaviside step function and $\gamma' (\en) \equiv 
\devo{\gamma}{\en} = (1 + 2 \alpha \en)$.
So doing, the expansion \eq{svp2} writes
\begin{displaymath}
\fk \simeq \frac{N(\en)}{\sigma(\en)} + 
\left(\frac{8}{3} \pi \frac{\sqrt{\mass}}{\hbar^3} \gamma(\en)\right)^{-1} 
P(\en)  \cos{\theta} \label{svp3} \p
\end{displaymath}
These new variables have also a direct physical interpretation:
\begin{equation}
\Ie{N(\en)} = \Ik{f_{0}(\en)}~~~,~~~\Ie{P(\en)} = \Ik{v(\en) f_{1}(\en)} \sv \label{signf}
\end{equation}
($ v(\en) =|{\bf v}_x|$) and are very suitable for our problem. \\
With these choices the equations of our SHE model, 
in the stationary homogeneous case, write:
\begin{eqnarray}
-\ecar E \devo{P}{\en} = G_{1}(N)  \label{eqdim1}\\
-\ecar E \devo{(g(\en) N)}{\en} +\ecar E h(\en) N = G_{2}(P) \label{eqdim2} 
\end{eqnarray}
where 
\begin{eqnarray}
G_{1}(N) &=& \Kbop \sigma(\en)
\left[(\nq+1) N(\en +\hw) + \nq N(\en-\hw)\right] - \nonumber \\
& &\Kbop\left[\nq\sigma(\en+\hw) -(\nq+1)\sigma(\en-\hw)\right] N(\en) \sv \nonumber \\
G_{2}(P) &=&
 -\left[\nq \Kbop \sigma(\en+\hw) + (\nq+1)  \Kbop \sigma(\en-\hw)+ \Kbac \sigma(\en) \right]
P(\en) \nonumber 
\end{eqnarray}
and
\begin{displaymath}
g(\en) \ud \frac{2}{3} \frac{\gamma(\en)}{\mass (\gamma'(\en))^2} ~~,~~
h(\en) \ud \frac{1}{\mass \gamma'(\en) } - \frac{4}{3} \frac{\alpha}{\mass}
\frac{\gamma(\en)}{(\gamma'(\en))^3} \p
\end{displaymath} 
We would like to underline that equations~\eq{eqdim1}-\eq{eqdim2} can be obtained
directly from the BTE by using a new alternative procedure. It consists 
in multiplying both sides of equation~\eq{BTE} respectively
by $\deet$ and by $\vk \deet$ and then formally 
integrating with respect to $\bk$  over the whole space $\Rk$. 
Some suitable closure
relations are needed: in particular by assuming that $f$ 
depends on $\bk$ only
through the variable $\en$ one obtains equations~\eq{eqdim1}-\eq{eqdim2} 
in the general non-homogeneous, non-stationary case 
(Liotta and Majorana,~1999; Majorana,~1998).
This method is similar to 
the method of frequency dependent moments of radiation hydrodynamics 
(Thorne,~1981).
%
\section{Dimensionless equations and physical parameters.}
It is useful to introduce dimensionless variables: let 
\begin{displaymath}
 t_{*} \ud \left[ 4 \sqrt{2} \pi 
 \left( \frac{\sqrt{ \mass} }{\hbar} \right)^{3} \sqrt{\kT} \Kbop \nq 
\right]^{-1} \sv \quad
\ell_{*} \ud \left({\frac{\kT}{\mass}}\right)^{\frac{1}{2}} t_{*} \sv \quad
\en_{*} \ud \kT \sv
\end{displaymath}
\begin{displaymath}
w \ud \frac{\en}{\en_{*}} \sv \quad 
n(w) \ud u_{*} \ell_{*}^{3} N(\en) \sv \quad
p(w) \ud u_{*} \ell_{*}^{2} t_{*} P(\en) \sv \quad
\end{displaymath}
\begin{displaymath}
\lambda \ud \frac{\hw}{\kT} \sv \quad
a \ud \frac{\nq + 1 }{\nq} = e^{\lambda} \sv \quad
\kappa \ud \frac{\Kbac}{\nq \Kbop} \sv \quad 
\zeta \ud \ecar E \frac{\ell_{*}}{u_{*}}  \sv \quad
\beta \ud \alpha~\en_{*} \p
\end{displaymath}
Moreover in the following we put $\chi (w) \ud w(1+\beta w)$ and $\chi'(w) \equiv 
\devo{\chi}{w} = ( 1 + 2 \beta w)$. 
By using these new variables, equations~\eq{eqdim1}-\eq{eqdim2} become
\begin{eqnarray}
-\zeta \devo{p}{w} = \mu(w) 
\left[a n(w+\lambda) + n(w-\lambda) \right] 
- \left[\mu(w+\lambda) + a \mu(w-\lambda)\right] n(w) \label{eqadim1} \\
\zeta \left[-\devo{(r(w)n)}{w} +q(w)n \right] = 
-\left[\mu(w+\lambda) +a \mu(w-\lambda) +\kappa \mu(w) \right] p(w) \label{eqadim2}
\end{eqnarray}
where
\begin{eqnarray}
\mu(w) & \ud & H(w) \left[\chi(w)\right]^{\frac{1}{2}} \chi'(w) \nonumber \\
r(w) & \ud & \frac{2}{3} \frac{\chi(w)}{\left[\chi'(w)\right]^2} \nonumber \\
q(w) & \ud & \frac{1}{\chi'(w)} - \frac{4}{3} \beta \frac{\chi(w)}{\left[\chi'(w)\right]^3} \p  \nonumber
\end{eqnarray}
We associate the following conditions to equations~\eq{eqadim1}-\eq{eqadim2}
\begin{eqnarray}
& & n(0) = 0 ~(\Rightarrow p(0) = 0)   \sv \qquad \lim_{w \freccia + \infty} p(w) = 0 \nonumber
\\
& & n(w) \geq 0 ~~\forall~w \geq 0 \sv \qquad \int_{0}^{+\infty} n(w) \, dw > 
0 ~~and ~~< + \infty \p \nonumber
\end{eqnarray}
Since equations~\eq{eqadim1}-\eq{eqadim2} are linear and homogeneous, 
if a solution  $(n(w), p(w))$, satisfying the above conditions exists, 
then, for every $c > 0$,  also $(c \hspace{1pt} n(w), c \hspace{1pt} p(w))$ 
is a solution. \\
The appropriate values of the physical parameters, in the case of a silicon device,
are given in table I, were $m_{e}$ denotes the electron rest mass.
\begin{center}
\begin{tabular}{|l|l|l|}
\hline
$ \mass = 0.32 \, m_{e}$ & $ \TL = 300 $ K & $\hw = 0.063$ eV \\[7pt]
$ \dm \Kbop = \frac{\left( D_{t} K \right)^{2}}{8 \pi^{2} \rho \omega_{op}} $ &
$ D_{t} K = 11.4 $ eV \mbox{$\stackrel{\circ}{\mbox{\rm A}}$}$^{-1}$ &
$\rho = 2330$ Kg m$^{-3}$ \\[15pt]
$ \dm \Kbac = \frac{\kT}{4 \pi^{2} \hbar v_{0}^{2} \rho} \Xi_{d}^{2} $ 
& $\Xi_{d} = 9$ eV & $v_{0} = 9040$ m sec$^{-1}$. \\[15pt]
$ \alpha = 0.5 \, eV^{-1}$ & $ \mbox{ } $ & $ \mbox{ } $  \\[7pt]
\hline
\end{tabular}
\end{center}
\begin{center}
{ \small {\bf Table I.}
Values of the physical parameters used in this paper. }
\end{center}
Using these parameters, we get $\lambda \simeq 2.437$, $\kappa \simeq 
5.986$ and $\beta \simeq 0.012926$.
\section{Asymptotic equations.}
Now we want to show that it is possible to find an
approximate solution of the equations \eq{eqadim1}-\eq{eqadim2} 
valid for high values of the electron energy. \\
It is useful to introduce a new variable $\psi(w)$ defined by
\begin{equation}
n(w) = \mu(w) \psi(w) \label{psidef} \p
\end{equation}
Equations~\eq{eqadim1}-\eq{eqadim2} become
\begin{eqnarray}
-\zeta \devo{p}{w} &=& \mu(w) 
\left[a \mu(w+\lambda) \psi(w+\lambda)+ \mu(w-\lambda) \psi(w-\lambda) \right] 
- \nonumber \\
& & \mu(w) \left[\mu(w+\lambda) + a \mu(w-\lambda)\right] \psi(w) 
\label{eqoas1} \\
\frac{2}{3} \zeta \frac{\left[ \chi(w)\right]^{\frac{3}{2}}}{\chi'(w)} 
\devo{\psi}{w} &=& 
-\left[\mu(w+\lambda) +a \mu(w-\lambda) +\kappa \mu(w) \right] p(w) 
\label{eqoas2} \p
\end{eqnarray}
Because we look for an asymptotic form of the equations \eq{eqoas1}-\eq{eqoas2}
for large values of the energy $w$, we expand the coefficients of the 
equations up to the zeroth order in $\lambda$: $\mu(w \pm \lambda) \simeq 
\mu(w)$. In this way we obtain a new set of equations
\begin{eqnarray}
-\zeta \devo{p_{A}}{w} &=& \mu^{2}(w)
\left[a \psi_{A}(w+\lambda) + \psi_{A}(w-\lambda) -(a+1)\psi_{A}(w) \right]
\label{eqas1} \\
p_{A}(w) &=& \frac{2}{3} \zeta \frac{\left[\chi(w)\right]^{\frac{3}{2}}}
{\chi'(w) \mu(w) \left[1+a+\kappa\right]} \devo{\psi_{A}}{w} 
\label{eqas2} \sv
\end{eqnarray}
where the subscript $A$ label the new unknowns. 
Substituting \eq{eqas2} into \eq{eqas1}, it follows 
\begin{eqnarray}
- \frac{2}{3} \frac{\zeta^2}{(1+a+\kappa)}
\left[ \frac{\chi(w)}{\left[\chi'(w)\right]^2} \psi_{A}'' +
\left(\frac{1}{\chi'(w)} -\frac{4 \beta \chi(w)}
{\left[\chi'(w)\right]^3}\right)\psi_{A}'\right] = \nonumber \\
\chi(w) \left[\chi'(w)\right]^{2}
\left(a \psi_{A} \left(w+\lambda \right) 
+ \psi_{A} \left(w-\lambda \right) -
(a+1) \psi_{A} \left( w \right) \right) \label{eqas3} \sv
\end{eqnarray}
where the primes denote derivatives with respect to $w$.
\section{Approximate solution.}
In order to find an approximate solution of equation~\eq{eqas3} we
expand the coefficients up to the first order in the 
non-parabolicity parameter $\beta$. 
\begin{eqnarray}
- \frac{2}{3} \frac{\zeta^2}{(1+a+\kappa)}
\left[ \left(w-3 \beta w^2 \right) \psi_{A}'' +
\left(1-6 \beta w\right) \psi_{A}'\right] = \nonumber \\
\left(w+5\beta w^2 \right)
\left(a \psi_{A} \left(w+\lambda \right) 
+ \psi_{A} \left(w-\lambda \right) -
(a+1) \psi_{A} \left( w \right) \right) \label{eqas4} \p
\end{eqnarray}
This choice is justified by
the smallness of $\beta$ and by the Kane equation itself,
which is of the first order in the non-parabolicity parameter.\\
We will search for solutions of equation~\eq{eqas4} having the form
\begin{equation}
\psi_{A}(w) = e^{f(w)}~~{\sv}~~with~~ 
f(w) \ud \eta_{0} w +\eta_{1} \beta w^2 \sv
\label{sol1}
\end{equation}
where $\eta_{0}$ and $\eta_{1}$ are functions of the applied
electric force $\zeta$.\\
It is useful to observe that $f(w \pm \lambda) =
f(w) + f(\pm\lambda) \pm 2\eta_{1} \beta \lambda w$.
Expanding the following quantities up to the first
order in $\beta$ :
\begin{eqnarray}
e^{\pm 2 \eta_{1} \beta \lambda w} &\simeq& 
1 \pm 2 \eta_{1} \beta \lambda w \sv \nonumber \\
e^{f(\pm\lambda)} &\simeq& e^{\pm \eta_{0} \lambda}
\left(1+\eta_{1} \lambda^{2} \beta \right) \sv \nonumber
\end{eqnarray}
substituting \eq{sol1} into \eq{eqas4}
and retaining only terms up to first 
order in $\beta$, we obtain, after dropping the common
factor $e^{f(w)}$, the equation
\begin{eqnarray}
& &- \frac{2}{3} \frac{\zeta^2}{(1+a+\kappa)}
\left[ \eta_{0} + \left(\eta_{0}^{2} + 4 \beta \eta_{1} -
6 \beta \eta_{0}\right)w +\left(-3\beta \eta_{0}^{2} +
4 \beta \eta_{0} \eta_{1} \right) w^2 \right] = \nonumber \\
& &~~~~ \left[ \left(a~e^{\eta_{0} \lambda} + e^{-\eta_{0} \lambda}\right)
\left(1+\beta \eta_{1} \lambda^2\right) -(a+1)\right]w + \nonumber \\
& &~~~~ \left[a~e^{\eta_{0} \lambda} 
\left(2 \eta_{1} \beta \lambda +5 \beta \right)
+ e^{-\eta_{0} \lambda}\left(-2 \eta_{1} \beta \lambda +5 \beta \right)
-5\beta(a+1)\right]w^2 \p \label{tra}
\end{eqnarray}
If we divide both sides of equation~\eq{tra} by $w^2$, and neglect the
terms in $\frac{1}{w^2}$, but not those in $\frac{1}{w}$
(in some sense we are serching for a "{\it weakly asymptotic}" solution), 
we obtain the following system of two transcendent 
equations in the unknowns $\eta_{0}$ and  $\eta_{1}$
\begin{eqnarray}
& &  \hspace{-12mm} - \frac{2}{3} \frac{\zeta^2}{(1+a+\kappa)}
\left(\eta_{0}^{2} - 6 \beta \eta_{0} + 4 \beta \eta_{1} \right) =
\left(a~e^{\eta_{0} \lambda} + e^{-\eta_{0} \lambda}\right)
\left(1+\beta \eta_{1} \lambda^2\right) -(a+1) \label{tra1} \\
& & \hspace{-12mm} - \frac{2}{3} \frac{\zeta^2}{(1+a+\kappa)}
\left(4 \eta_{0} \eta_{1} -3 \eta_{0}^{2}\right) =
a~e^{\eta_{0} \lambda}
\left(5+ 2 \eta_{1} \lambda \right)
+ e^{-\eta_{0} \lambda}\left(5 - 2 \eta_{1} \lambda \right)
-5(a+1) \sv \label{tra2}
\end{eqnarray}
where in the second equation we have dropped the common factor $\beta$.
If we are able to solve the previous system, it is possible to 
obtain $\eta{_0}$ and $\eta{_1}$ as functions of the applied
electric force $\zeta$, and then $\psi_{A}(w)$. 
Moreover, by using equation~\eq{eqas2}, one can find $p_{A}(w)$.\\
Henceforth we will indicate as {\it asymptotic solution of the SHE
equations} the expressions of $n_{A}$ anp $p_{A}$ which can be obtained by
means of the approximate solution of \eq{eqas1}-\eq{eqas2} which have been found.
\section{Discussion of the solution and comparison with numerical 
results.}
An analytical solution of the 
equations \eq{tra1}-\eq{tra2} has turned out to be impossible.
Therefore we have limited ourselves to a graphical and numerical analysis.
Given  a value of the applied electric force $\zeta$, the requirement
that the functions be integrable in $[0,+\infty[$ tells us that both
$\eta_{0}$ and $\eta_{1}$ have to assume negative values.  
In fact, we found a negative solution of the system \eq{tra1}-\eq{tra2}
in the domain $[-1,0] \times [-1,0]$ of the $(\eta_{0},\eta_{1})$ plane,
for all the values of the electric field in the explored range.
We used a simple MapleV algorithm in order to find the solutions. 
In table II we give some of the values we found.
\begin{center}
\begin{tabular}{|l|l|l|}
\hline
$E$ (V/cm) & $ ~~~\eta_{0}$ & $~~~\eta_{1}$  \\
\hline
$0.0$ &  $ -1.0 $   &   $ ~~0.0$  \\
$1.0 \times 10^{2}$ &  $ -0.9999636274$   &   $-0.0001493648$  \\
$1.0 \times 10^{3}$ &  $ -0.9963693066$   &   $-0.0148712049$  \\
$5.0 \times 10^{3}$ &  $ -0.9136708716$   &   $-0.3294754506$  \\
$1.0 \times 10^{4}$ &  $ -0.7162321385$   &   $-0.8333494338$  \\
$3.0 \times 10^{4}$ &  $ -0.2511364699$   &   $-0.5681222515$  \\
$5.0 \times 10^{4}$ &  $ -0.1270766899$   &   $-0.2772663271$  \\
$7.0 \times 10^{4}$ &  $ -0.0780599266$   &   $-0.1591756496$\\
$1.0 \times 10^{5}$ &  $ -0.0452815087$   &   $-0.0850850173$  \\
\hline
\end{tabular}
\end{center}
\begin{center}
{ \small {\bf Table II.} Some values of $\eta_{0}$ and $\eta_{1}$ 
as functions of the electric field. }
\end{center}
In figures 1, 2 and 3, we compare the asymptotic
solution $(n_{A},p_{A})$  and the numerical solution $(n_{N},p_{N})$
of equations~\eq{eqadim1}-\eq{eqadim2}, the latter obtained by a
suitable numerical technique (Liotta and Majorana,~1999), 
for some values of the applied electric field.
\begin{figure*}[b!]
\centerline{\psfig{figure=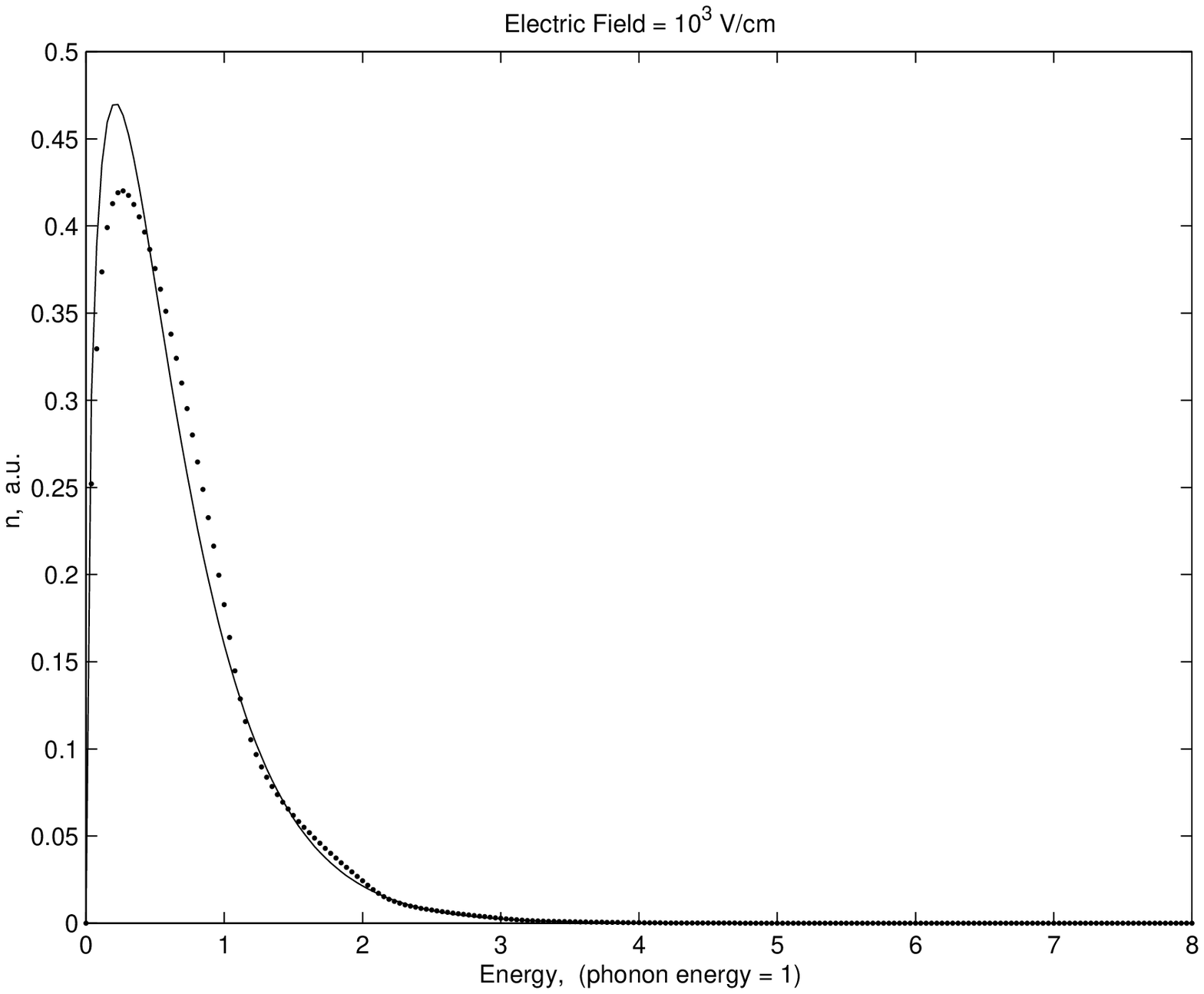, height=4.9 cm}\psfig{figure=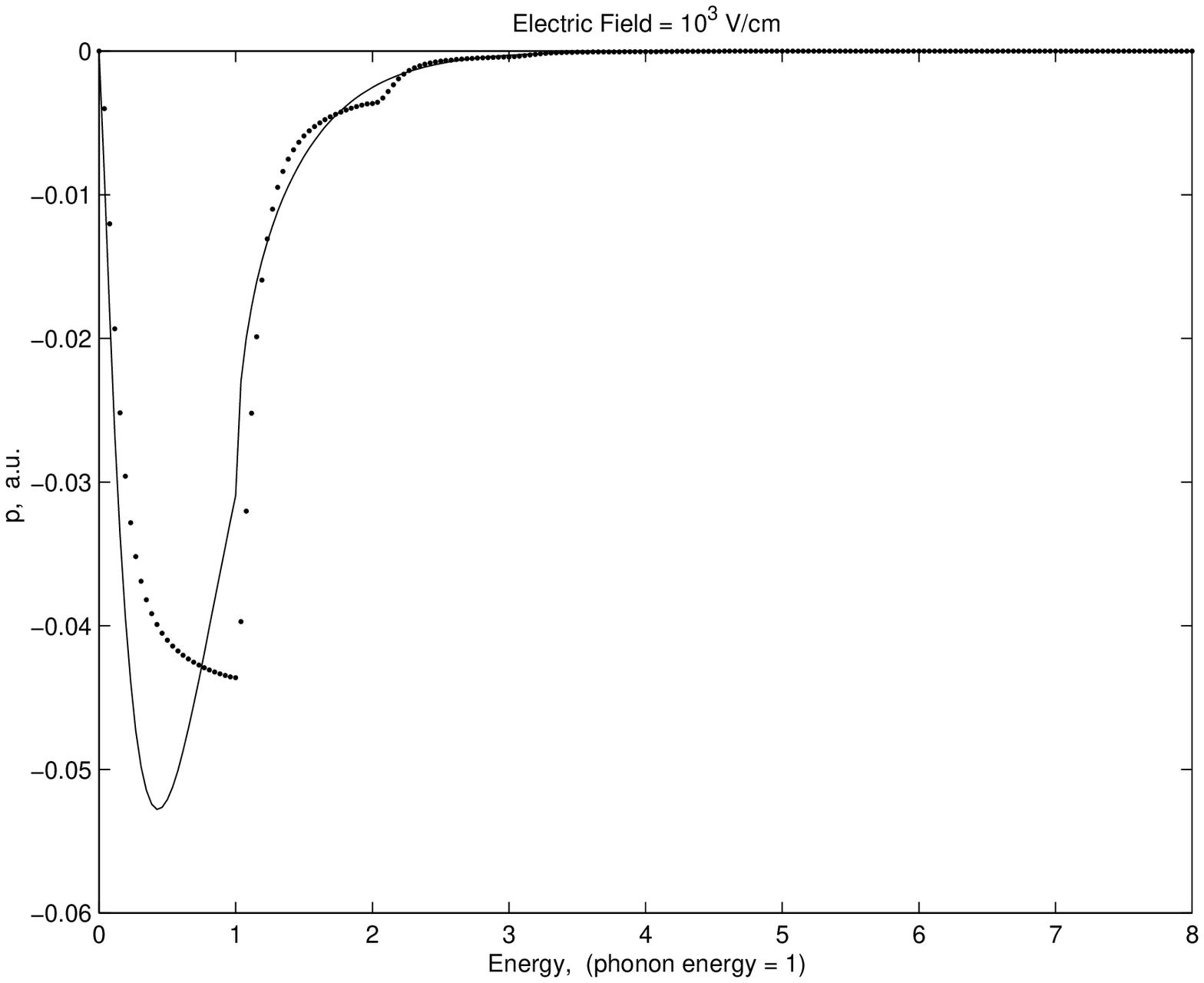, height=4.9 cm}}
\vspace{-0.15 in}
\caption{\small  Electric field =  $10^3$ V/cm. $n$ and $p$ versus energy.
Dots: numerical solution, continuous line: asymptotic solution. The optical phonon
energy is set equal to one.}
\end{figure*}
\begin{figure*}[h!]
\vspace{-0.1in}
\centerline{\psfig{figure=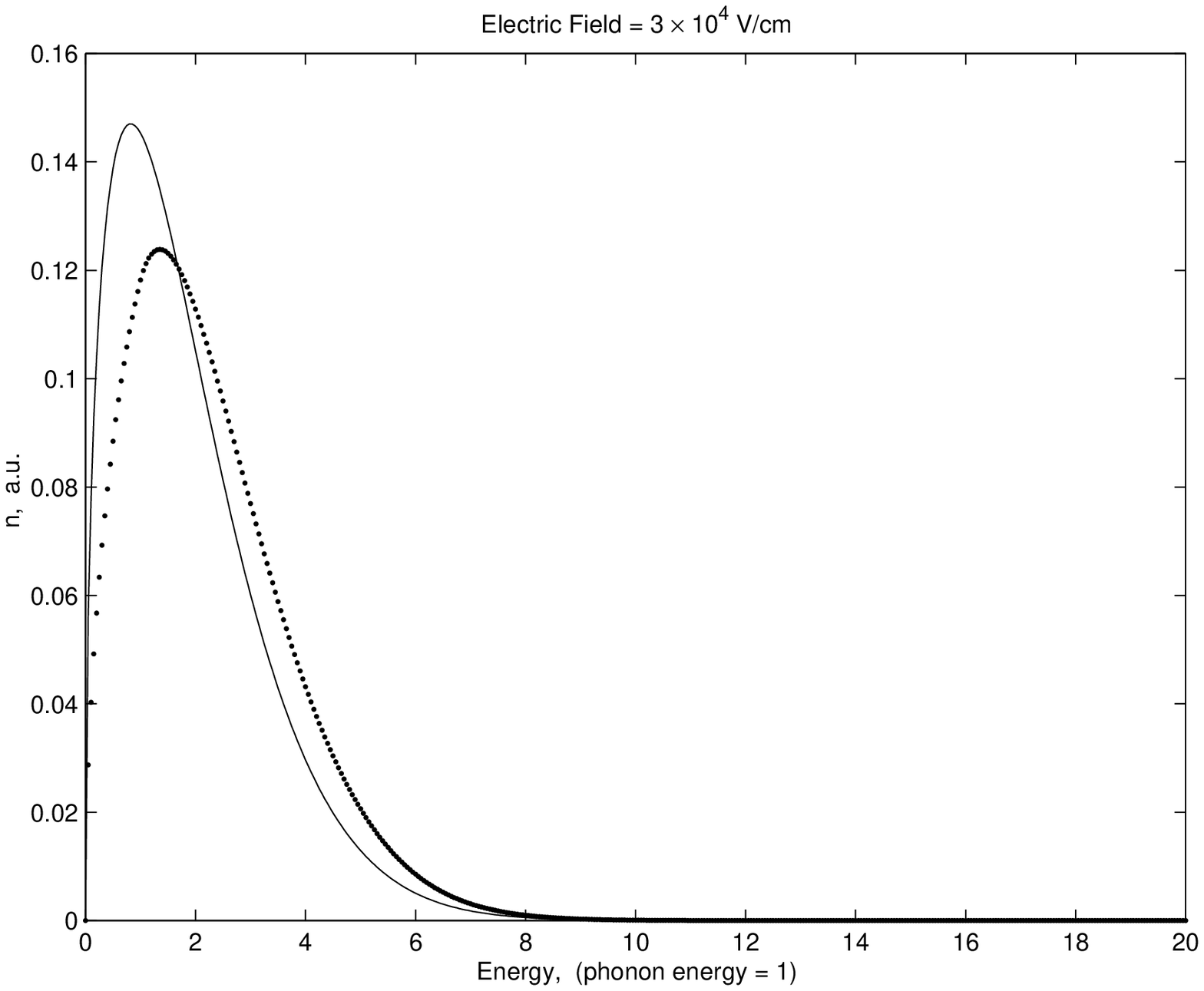, height=4.9 cm}\psfig{figure=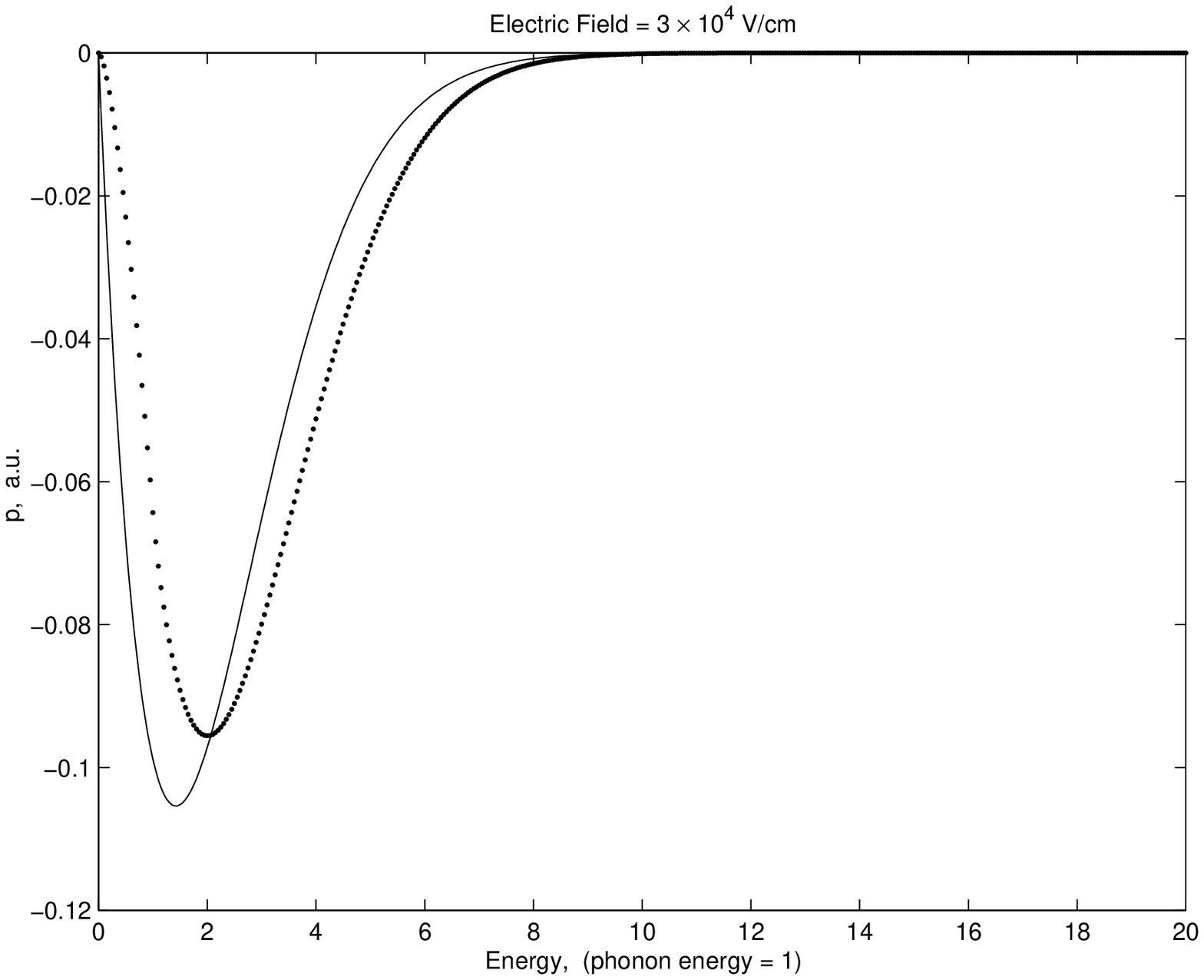, height=4.9 cm}}
\vspace{-0.15 in}
\caption{\small  Electric field = $3 \times 10^4$ V/cm. }
\vspace{-0.1 in}
\end{figure*}
\begin{figure*}[h!]
\centerline{\psfig{figure=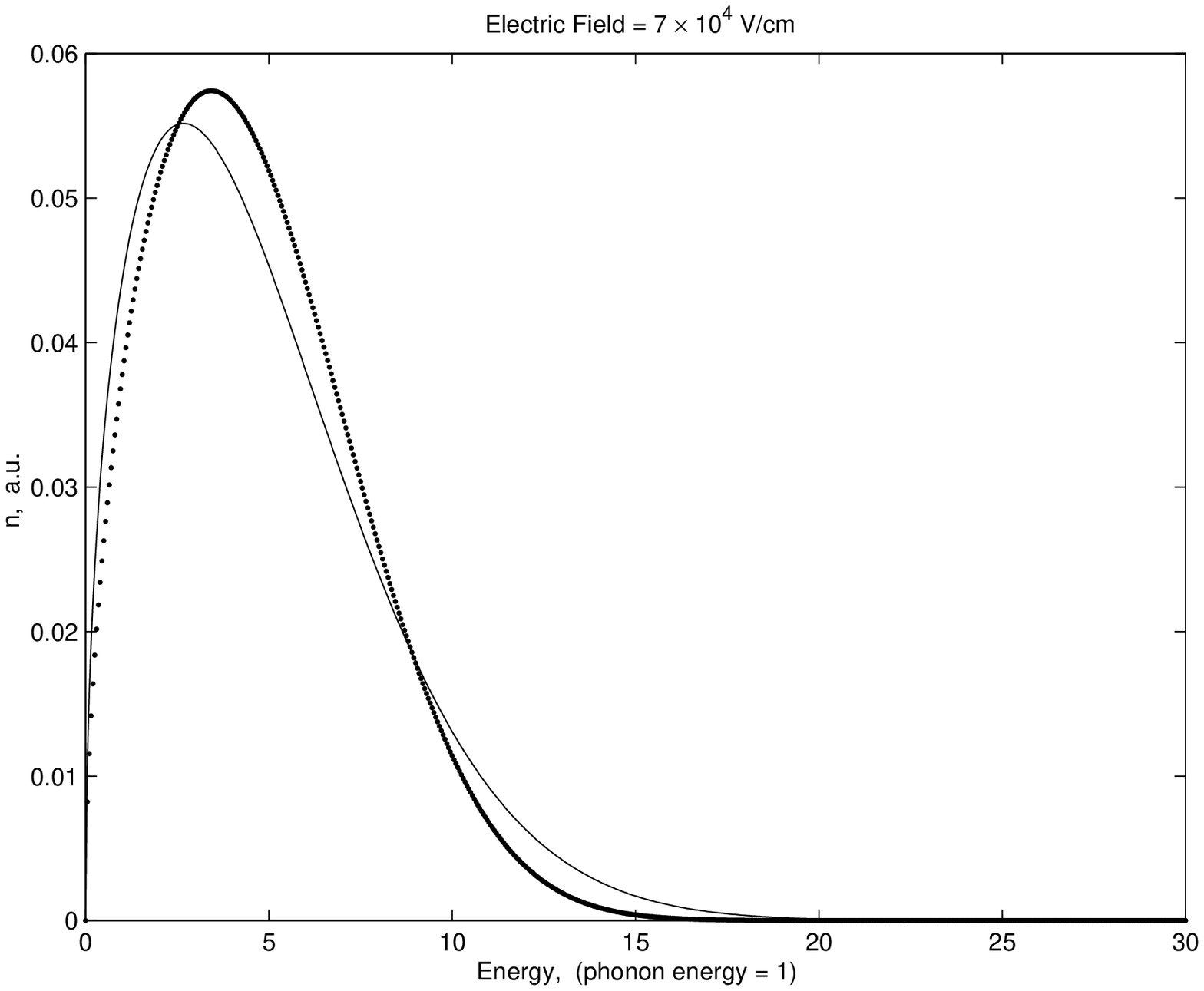, height=4.9 cm}\psfig{figure=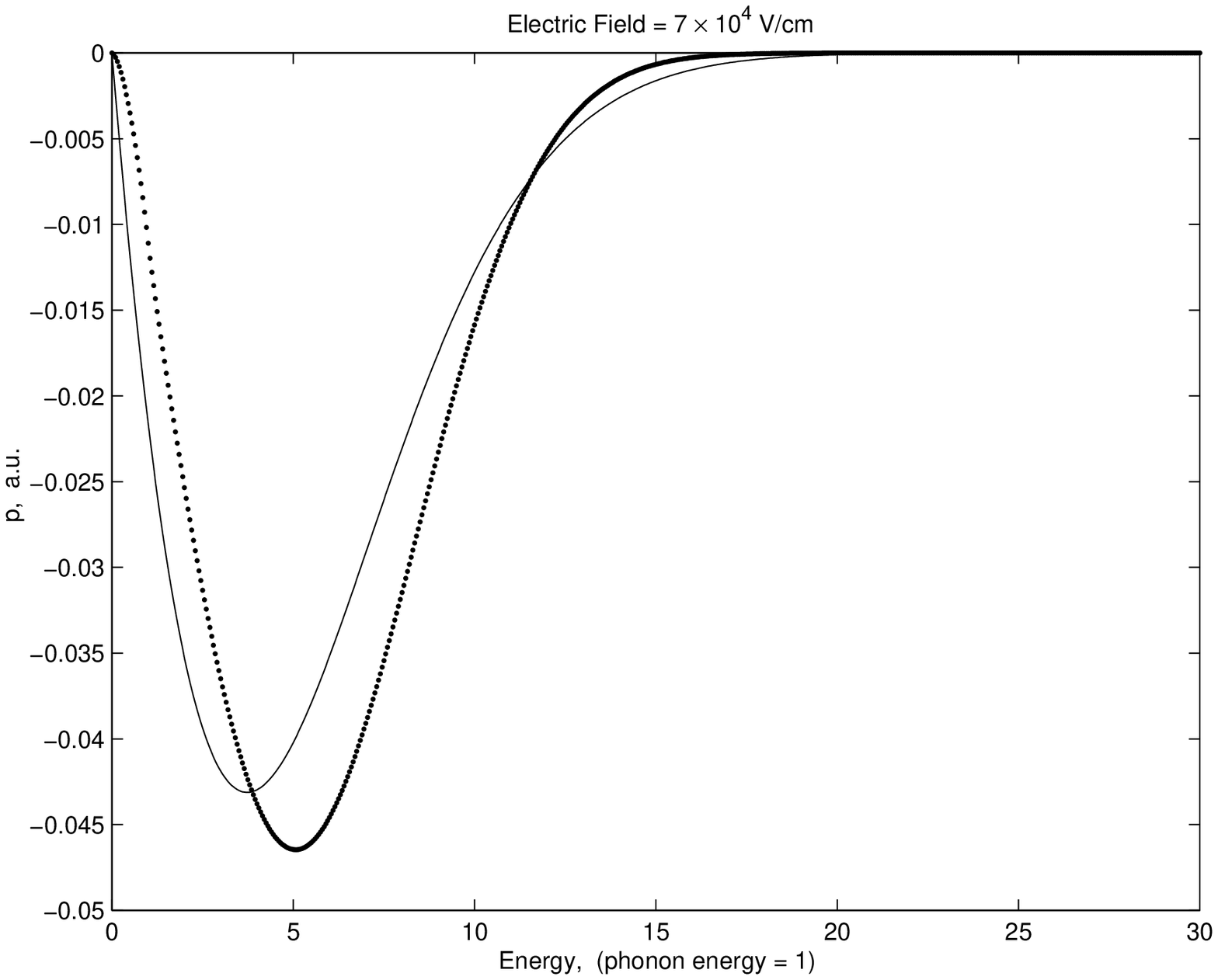, height=4.9 cm}}
\vspace{-0.15 in}
\caption{\small Electric field = $7 \times 10^4$ V/cm.}
\vspace{-0.1 in}
\end{figure*}
%
As is possible to see, the agreement between the numerical and asymptotic 
solutions is good in all the energy range, despite the obtained asymptotic 
solution should be good only for high enough energy values
(the agreement is obviously very good in this region). 
We want to observe that for low values of the electric field 
($| \bE | \leq 10^3$ V/cm), 
instead of formula \eq{eqas2}, we use the following expression for $p_{A}$:
\begin{equation}
p_{A}(w) =\frac{1}{\mu(w+\lambda) +a \mu(w-\lambda) + \kappa \mu(w)} 
\frac{2}{3} \zeta \frac{\left[\chi(w)\right]^{\frac{3}{2}}}
{\chi'(w)} \devo{\psi_{A}}{w} 
\label{eqas2mod} \p
\end{equation}
This choice is  not coherent with the expansion, but allows a better
agreement with the numerical solution and puts in evidence a discontinuity
in the derivatives at the point $w=\lambda$ 
(in dimensional variables $\en = \hw$) due to the term
$\mu(w-\lambda)$, that is zero for $0 \leq w \leq \lambda$. 
This term is also present in the original set of equations~\eq{eqadim1}-\eq{eqadim2}.
Then, such discontinuity is also expected in the solutions.\\
As a measure of the difference between the numerical and asymptotic solution 
we can also compute $V_{as}=\int p_{A} d \en/ \int n_{A} d \en$ 
and $V_{num}=\int p_{N} d \en/ \int n_{N} d \en$, which 
give, in dimensional units, the electron hydrodynamical velocity. In 
table III  we compare the values we found.
\begin{center}
\begin{tabular}{|l|l|l|}
\hline
$E$ (V/cm) & $ ~~~V_{as} (m/sec)$ & $~~~V_{num} (m/sec)$  \\
\hline
$1.0 \times 10^{2}$ &  $ 1.4879 \times 10^{3}$   &   $1.4537 \times 10^{3}$  \\
$1.0 \times 10^{3}$ &  $ 1.4831 \times 10^{4}$   &   $1.3858 \times 10^{4}$  \\
$5.0 \times 10^{3}$ &  $ 7.0223 \times 10^{4}$   &   $5.1060 \times 10^{4}$  \\
$1.0 \times 10^{4}$ &  $ 6.0652 \times 10^{4}$   &   $7.3611 \times 10^{4}$  \\
$3.0 \times 10^{4}$ &  $ 9.8325 \times 10^{4}$   &   $9.7714 \times 10^{4}$  \\
$5.0 \times 10^{4}$ &  $ 1.0106 \times 10^{5}$   &   $9.9872 \times 10^{4}$  \\
$7.0 \times 10^{4}$ &  $ 9.5124 \times 10^{4}$   &   $9.8395 \times 10^{4}$  \\
$1.0 \times 10^{5}$ &  $ 8.2528 \times 10^{4}$   &   $9.5101 \times 10^{4}$  \\
\hline
\end{tabular}
\end{center}
\begin{center}
{ \small {\bf Table III.} Comparison between the electron hydrodynamical velocities
calculated by using respectively the asymptotic and numerical solutions. }  
\end{center}
The behaviour of the hydrodynamical velocity is the typical one 
when the Kane equation is used (see Tomizawa,~1993, p. 100, fig. 3.11).
We do not consider higher values of the electric field because 
in this case also the Kane equation becomes inadequate to describe 
the conduction band, and a full band structure should be used
(Vecchi and Rudan,~1998).
%
\section{Conclusions and acknowledgments.}
We have shown that it is possible to find an "{\it asymptotic solution}" 
of the SHE equation \eq{eqadim1}-\eq{eqadim2} in which the dependence on 
the applied electric field is given implicitly through the system of 
transcendent equations \eq{tra1}-\eq{tra2}.
The solution of the system \eq{tra1}-\eq{tra2} can be obtained by 
using standard numerical techniques. The agreement with the numerical 
solution is good in all the explored range of applied electric fields.\\
If we put $\beta=0$, we recover the parabolic band case and the asymptotic 
solution reduces to the asymptotic one already found 
by Liotta and Majorana~(1999)
(L-M solution). 
The L-M solution was used by Anile and Mascali~(2000) 
to obtain
a two fluid hydrodynamical model, where the electron population is 
splitted in two sub-populations: low-energy and high-energy electrons, 
separated by a threshold energy. 
In order to close the moment equations relatively to the hot electrons
they used a distribution function whose form is directly suggested by the 
L-M solution, whereas relatively to cold electrons a 
a standard maximum-entropy distribution function is utilized. 
A detailed study of the resulting
high-energy hydrodynamical model is given in Anile, Liotta and Mascali~(2000).
Some works about the
extension of this model to the case of Kane band, using the asymptotic 
solution found in this paper, are in progress. 
A better description of the hot electron population
is expected. \\
The author acknowledges support from Italian CNR
 (Prog.~N.96.03855.CT01), from TMR  (Progr. n. ERBFMRXCT970157),
from Italian MURST (Prot. n. 9801169828-005) and from 
"{\it Convenzione quadro Universit\`{a} di Catania-ST 
Microelectronics}".\\
The author wish also to thank prof. A.~M.~Anile, prof. A.~Majorana
and dr. G. Mascali for useful discussions and precious suggestions. 

\end{document}